# Determination of Poynting Vector Characteristics

I. Mokhun,[1,*], V. Danko,[2,†] A. Kovalenko,[2,†]

[1] Yuriy Fedkovych Chernivtsi National University, 2, Kotsiubynskoho St., Chernivtsi 58002, Ukraine

[2] Taras Shevchenko National University of Kyiv, 64/13, Volodymyrska St., Kyiv, 01601, Ukraine

[†]These authors contributed equally to this work.

* i.mokhun@chnu.edu.ua

**This paper presents a novel method for measuring the Poynting vector characteristics of monochromatic electromagnetic waves. We outline a specific design for such a meter and provide experimental data to validate the approach. For testing purposes, we utilized vortex beams with both linear and circular polarization.**

## 1. INTRODUCTION

Recently, there has been a surge of interest in singular optics, a relatively new field within classical photonics (see, for example, [1–4]). This interest stems primarily from the fact that singular optics provides a fresh perspective on established optical principles, deepening our understanding of known phenomena and often revealing new insights into the propagation and transformation of electromagnetic waves. This is largely due to a modern focus on the structure of energy flows [1–7] and their associated angular momentum (AM).

While interest in this area did not emerge exclusively with the "era of singular optics" – as evidenced by early studies on angular momentum in the last century [8,9]. The field reached a turning point with the work published in [10]. By formalizing the concepts of orbital (OAM) and spin (SAM) angular momentum, that study catalyzed the rapid development of optical tweezers, a key applied branch of modern optics (see, for example, [11,12]).

Furthermore, angular momentum can be calculated relative to any point in the field. In practice, this quantity is typically "tethered" to a specific origin, often a Poynting singularity [4,13,14] – a point where the magnitude of the Poynting vector vanishes. Consequently, characterizing the behavior of the Poynting vector, which governs "energy flows" [1,4–7], enables the immediate analysis of the AM distribution across any region of the field.

It is also worth noting that analyzing the transformation of the Poynting vector's characteristics is of intrinsic interest. For instance, it has been demonstrated that during diffraction at an opaque edge, the energy structure of uniformly elliptically polarized beams transforms in a manner analogous to an optical vortex, even while the intensity distributions remain similar to those of a scalar wave [15].

The energy formalism of electromagnetic wave propagation and transformation involves two facets: an analytical aspect, which is largely established, as Poynting vector characteristics can be derived from the field parameters and an experimental aspect. The latter requires the ability to directly measure the vector's characteristics, a capability that is crucial for practical applications.

It should be noted that direct measurement of Poynting vector characteristics specifically those transverse to the preferred direction of propagation remains a significant challenge. Now, researchers often rely on indirect methods that estimate the Poynting vector or AM through proxy quantities. For instance, the rotational velocity of a particle captured in an optical trap is frequently used to infer AM values [2,9,11,12]. Perhaps the most relevant approach was demonstrated in [16], which showed that large dielectric particles with near-dipole Mie scattering modes can, under specific conditions, experience a force directly proportional to the local Poynting vector.

However, these methods suffer from several inherent limitations:

- Medium interference: Measuring energy flux distributions across a beam's cross-section requires a suspension of test particles in a medium that must be perfectly transparent. Any absorption leads to convection currents from localized heating, which distorts measurements.
- Spatial uncertainty: Such measurements are impossible without the presence of a medium or additional elements, like a cuvette. Furthermore, particles are typically distributed randomly in both transverse and longitudinal directions. This stochastic distribution prevents high-precision mapping of the Poynting vector in a single, well-defined cross-sectional plane.

While the situation simplifies under the paraxial approximation, where the transverse Poynting vector can be reconstructed using full phasemetry of orthogonal components and local Stokes polarimetry [17]. At the same time this approach introduces its own set of drawbacks:

- Equipment complexity: It requires interferometric measurements and a coherent reference wave, which is not always feasible to generate.

- Stability requirements: The setup demands rigorous vibration isolation.
- Computational overhead: Extensive intermediate calculations are required to analyze the experimental data, significantly increasing the time necessary to recover the Poynting vector characteristics.

In this paper, we propose a modified method that bypasses the need for interferometry of orthogonal components, thereby eliminating these conventional limitations.

## 2. PRINCIPLES FOR DETERMINING POYNTING VECTOR CHARACTERISTICS

The principles of the proposed method are based on the following considerations. It is established that within the paraxial approximation, one possible representation of the transverse components of the Poynting vector is given by [4]:

$$\begin{cases} \bar{P}_x = \bar{P}_{x\,orb} + \bar{P}_{x\,sp} \\ \bar{P}_y = \bar{P}_{y\,orb} + \bar{P}_{y\,sp} \end{cases}, \tag{1}$$

where $\bar{P}_{i\,orb}, \bar{P}_{i\,sp}$, $i = x, y$ represent the orbital and spin components of the Poynting vector [5,18]. These components can be expressed by the following relations:

$$\begin{cases} \bar{P}_{x\,orb} = \frac{c\lambda}{16\pi^2}\left(I_x \frac{\partial\Phi_x}{\partial x} + I_y \frac{\partial\Phi_y}{\partial x}\right) \\ \bar{P}_{y\,orb} = \frac{c\lambda}{16\pi^2}\left(I_x \frac{\partial\Phi_x}{\partial y} + I_y \frac{\partial\Phi_y}{\partial y}\right) \end{cases}, \tag{2}$$

$$\begin{cases} \bar{P}_{x\,sp} = -\frac{c\lambda}{32\pi^2}\frac{\partial s_3}{\partial y} \\ \bar{P}_{y\,orb} = \frac{c\lambda}{32\pi^2}\frac{\partial s_3}{\partial x} \end{cases}, \tag{3}$$

where $I_i, \Phi_i, i = x, y$ denote the intensities and phases of the orthogonal electric field components, respectively, $s_3$ represents the fourth Stokes parameter, $c$ is the speed of light, and $\lambda$ is the wavelength.

An analysis of relations (1–3) demonstrates that if the measured values for $I_i, \frac{\partial\Phi_i}{\partial l}, \frac{\partial s_3}{\partial l}$, $i = x, y$, $l = x, y$ are obtained, the transverse components of the Poynting vector, $\bar{P}_x$ and $\bar{P}_y$, can be reconstructed.

This concept served as the foundation for the method described in [17], where the spatial distribution of phase derivatives was calculated by analyzing the phase profiles obtained through orthogonal component interferometry. However, interpreting these interferograms is often a formidable task, particularly when analyzing complex, non-trivial fields. Moreover, as previously noted, generating a stable reference beam is not always feasible.

Naturally, the local phase derivatives of orthogonal components can be determined via alternative methods that do not require a reference beam, most notably, the Hartmann method [19]. The principle of this technique is established as follows: a microlens array is positioned within the path of the scalar (linearly polarized) wave under investigation (see Fig. 1).

We assume that within the entrance pupil $d$ of each microlens, the intensity of the wave remains constant while the phase varies linearly. Under these conditions, the portion of the wavefront sampled by the microlens acts as a local plane wave with a specific intensity and tilt. This tilt is determined by the phase derivatives $\frac{\partial\Phi}{\partial x}, \frac{\partial\Phi}{\partial y}$ at the field point corresponding to the center of the microlens. In the case when the paraxial approximation is satisfied (that is realized practically always for each microlens) the phase derivatives at this point can be directly determined from the displacement of the focal spot $\Delta l$ in the focal plane:

$$\frac{\partial\Phi}{\partial l} = k\frac{\Delta l}{f}, \tag{4}$$

where $k = \frac{2\pi}{\lambda}$ is the wave number, $l = x, y$.

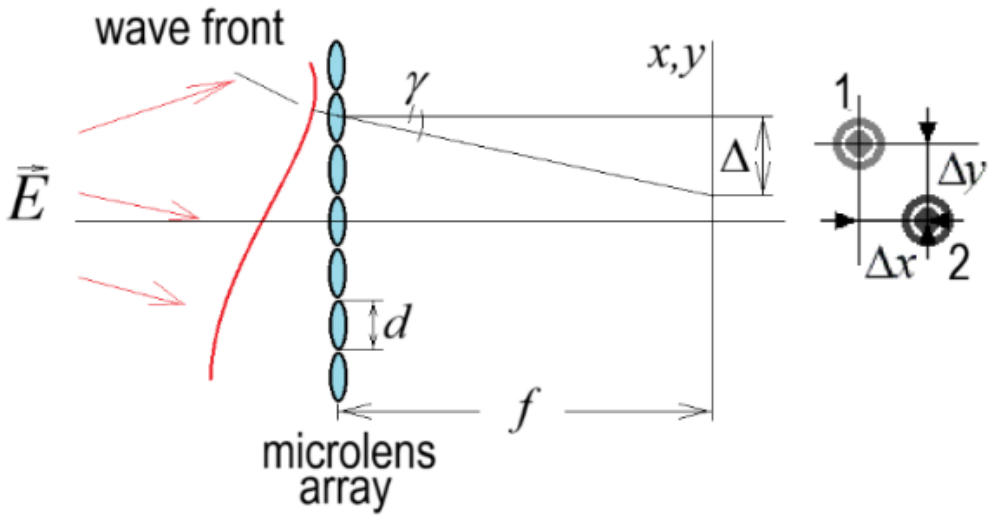

**Fig. 1.** Operating principle of the Hartmann method.
1 – focal spot position in the absence of wavefront curvature (reference position).
2 – displaced focal spot position, indicating the local wavefront tilt $\gamma$.

The peak intensity of the displaced focal spot is proportional to the local intensity of the field at the microlens center. The spatial resolution of this method is determined by the dimensions of the microlenses and the pitch (the distance between them) of the array.

Consequently, one possible experimental configuration for measuring the transverse components of the Poynting vector is illustrated in the schematic shown in Figure 2.

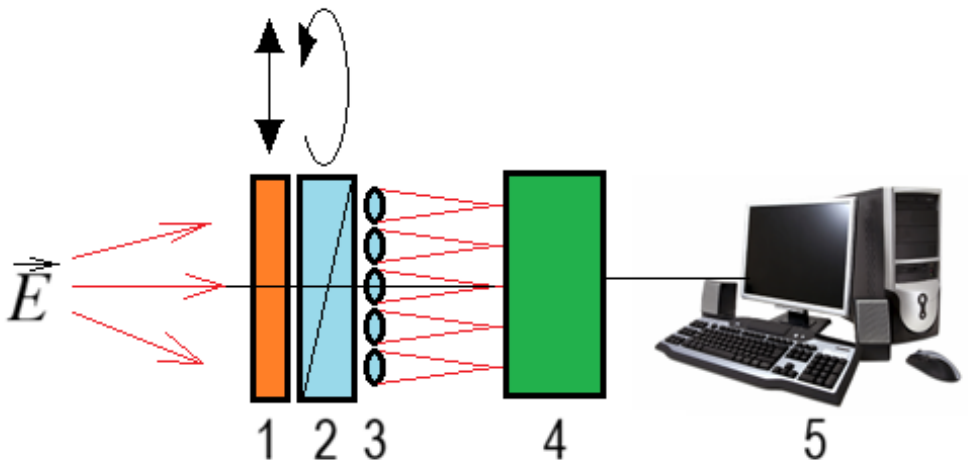

**Fig. 2.** Experimental schematic for reconstructing Poynting vector characteristics.
1 – quarter-wave plate; 2 – polarizer; 3 – microlens array; 4 – CCD camera; 5 – computer.

The incident field $\vec{E}$ first passes through the polarizer (2) in the absence of the quarter-wave plate (1) to separate one of the orthogonal components. Behind the polarizer, a microlens array (3) samples the field. In the focal plane, the system records both the intensity of this component and the focal spot displacements (the Hartmannogram), providing the data necessary to determine the spatial distribution of the local phase derivatives.

The polarizer axis is then rotated by 90° to perform identical measurements for the second orthogonal component. Subsequently, the polarizer is oriented at ±45°, and the quarter-wave plate (1) is introduced and appropriately aligned. This configuration functions as a circular analyzer, transmitting either right- or left-handed circular polarization to determine the spatial distribution of the fourth Stokes parameter, $s_3$ (see, for example, [20]). Finally, using these experimental datasets and relations (1–3), the characteristics of the transverse Poynting vector components are reconstructed.

## 3. EXPERIMENTAL VALIDATION: MEASURING POYNTING VECTOR COMPONENTS

Experimental validation of the proposed method was conducted using linearly and circularly polarized optical vortex beams, as the behavior of their transverse Poynting vector components is well-documented (see, for example, [2,4]).

The testing was performed using the setup illustrated in Fig. 3, which is an implementation of the configuration shown in Fig. 2. A quasi-Gaussian, linearly polarized beam from a He-Ne laser (1) was converted into a vortex beam using a computer-generated hologram (CGH) (2) [21,22].
In the −1st diffraction order of the hologram, a vortex with a topological charge of $m = +1$, was generated. Two experimental scenarios were then investigated:

- Scenario 1: The linearly polarized vortex was converted into a left- or right-handed circularly polarized state using an appropriately oriented quarter-wave plate (3).
- Scenario 2: In the absence of the plate, the vortex remained linearly polarized.

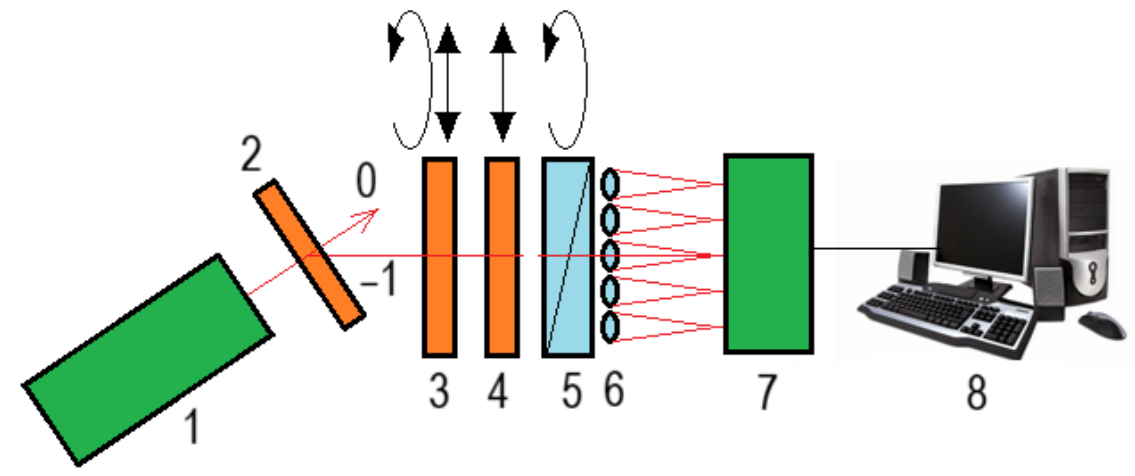

**Fig. 3.** Optical arrangement for experimental validation.
1 – He-Ne laser; 2 – computer-generated vortex hologram; 3,4 – quarter-wave plates; 5 – polarizer; 6 – microlens array; 7 – CCD camera; 8 – computer.
The beam subsequently passed through the analyzer comprising a second quarter-wave plate (4) and a polarizer (5) before being sampled by a microlens array (6). The resulting focal spots were captured by the CCD camera (7). The microlens array consisted of an 11x11 grid, with each microlens having a diameter of 0.4 mm and a focal length of 24 mm.

Figure 4 presents the reconstructed transverse Poynting vector characteristics for three distinct cases. Figure 4a illustrates the energy flow near the center of a linearly polarized vortex. Figure 4b depicts the behavior of a circularly polarized vortex where the sign of the topological charge ($m = +1$) and the handedness factor ($h = -1$, left-hand circular polarization) differ [4]. In this configuration, the OAM and the SAM have the same sign, and the orbital and spin components of the Poynting vector (as defined in relations 1–3) are unidirectional. Consequently, the resulting vector circulates around the beam axis with a magnitude approximately twice that of the linearly polarized case. A Poynting singularity [4] is clearly observed at the beam center, where the vector azimuth is indeterminate.

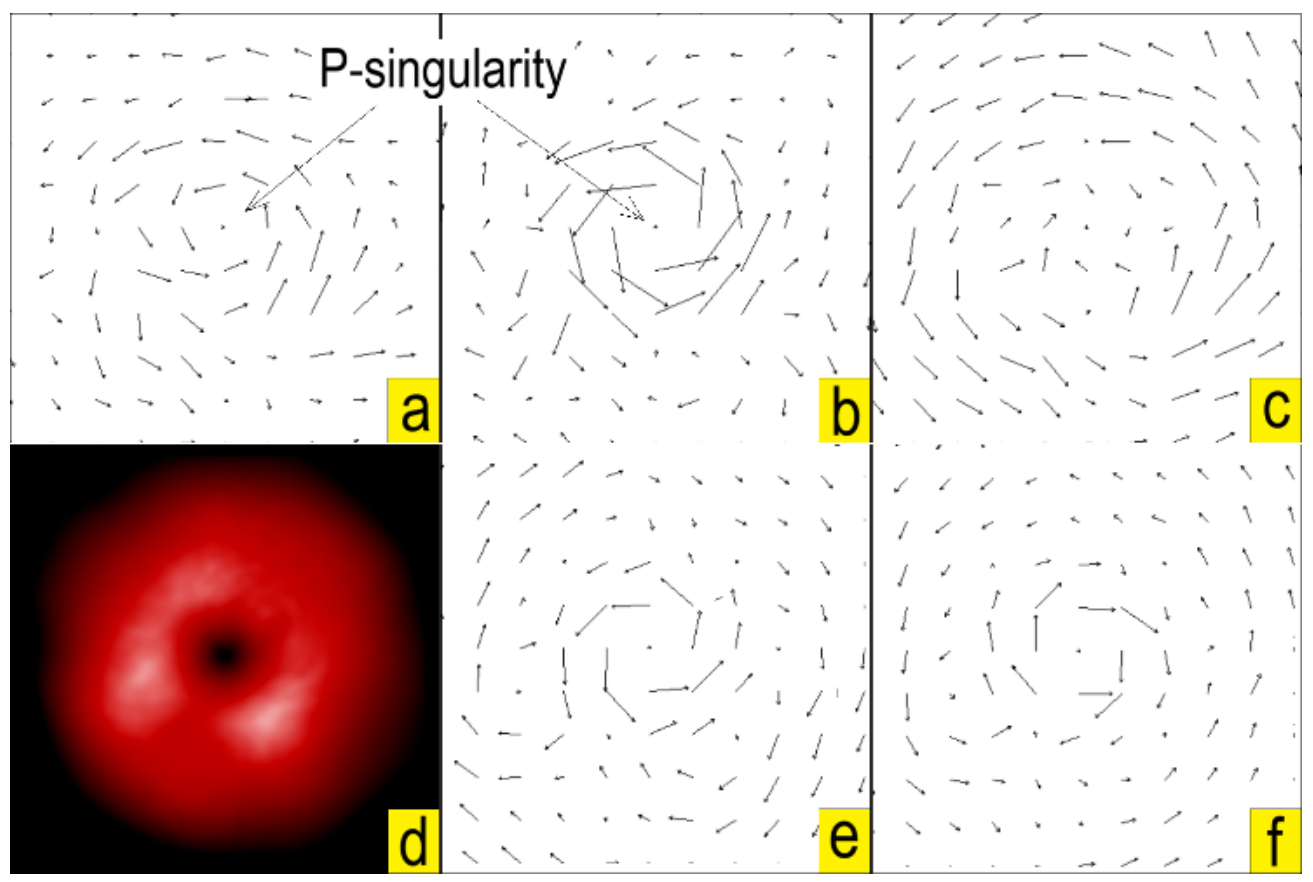

**Fig. 4.** Reconstruction of the transverse Poynting vector components for linearly and circularly polarized vortices.
(a) Reconstructed Poynting vector distribution in the vicinity of a linearly polarized vortex center. The arrow length represents the magnitude of the transverse component, while the arrow direction indicates the local azimuth. (b) Reconstructed Poynting vector near the center of a circularly polarized vortex where the signs of the topological charge and handedness factor are opposite (additive case). (c) Reconstructed Poynting vector where the signs of the charge and handedness factor are identical (subtractive case). (d) Measured intensity distribution of a linearly polarized vortex. (e) Behavior of spin energy fluxes when the signs of the charge and handedness factor are opposite. (f) Behavior of spin energy fluxes when the signs of the charge and handedness factor are identical.

Finally, Figure 4c illustrates the vector behavior when the signs of the topological charge and the handedness factor are the same ($m = +1, h = +1$). In this case, the orbital and spin components are directed antiparallel, leading to the near-total cancellation of the transverse Poynting vector throughout the central region of the vortex.

## 4. ANALYSIS, DISCUSSION, AND CONCLUSIONS

As expected, when the OAM and SAM of a circularly polarized beam have the same sign (where angular momenta magnitudes are additive), the transverse component of the Poynting vector circulates around the vortex center (Fig. 4b). The magnitude of the Poynting vector components for this circularly polarized state is approximately double that of a linearly polarized vortex. In both instances, a Poynting singularity is formed at the beam axis, where the transverse component vanishes and its azimuth remains undefined. We observe minor irregularities and deviations from perfect circulation in the vector's behavior (Fig. 4a), which we attribute to inherent experimental noise and measurement errors.

In the case where SAM compensates for OAM occurring when the topological charge and the handedness factor share the same sign (Fig. 4c) the transverse component of the Poynting vector becomes effectively zero across the central region of the vortex. The resulting chaotic vector orientation in this region is a natural consequence of the subtraction of two nearly equal values, where the reconstructed magnitude falls within the margin of experimental error.

Figure 4d illustrates the measured intensity distribution in the cross-section of the linearly polarized beam. Figures 4e and 4f illustrate the behavior of the spin component of the Poynting vector. As shown, the spin component circulates around the beam center in a manner similar to the orbital component of a linearly polarized vortex (Fig. 4a). However, when the signs of the charge and the handedness factor are identical, the circulation directions of the orbital and spin components are different, leading to the observed compensation and the near-disappearance of the vector at the beam center (Fig. 4c).

It should be noted that the microlens array parameters used in these test experiments were not optimized for this specific application. Nevertheless, our results demonstrate the clear viability of this method for measuring Poynting vector components. The spatial resolution and the fidelity of the vector reconstruction are directly governed by the microlens aperture and the array pitch.

Consequently, this method is most applicable to the analysis of fields with "slowly" varying structures. Specifically, the field correlation radius should ideally be significantly larger than the combined dimension of the microlens entrance pupil and the inter-lens distance. Furthermore, since the Hartmann method is fundamentally limited to waves with relatively small wavefront inclinations, the method is most effective for fields with "smooth" phase variations.

Despite these constraints, the avoidance of interferometric measurements in favor of the Hartmann method significantly simplifies the reconstruction of Poynting vector characteristics. This approach drastically reduces the required computational time, offering the potential for real-time implementation. Such a capability is particularly vital for practical applications, such as the dynamic control of particles in optical tweezers and the development of advanced micromanipulation systems.

**Funding.** This work was supported by the National Research Foundation of Ukraine under grant № 2025.06/0086.

**Disclosures.** The authors declare no conflicts of interest.

**Data availability.** Data underlying the results presented in this paper are not publicly available at this time but may be obtained from the authors upon reasonable request.

## REFERENCES

1. G.J. Gbur “Singular Optics” Taylor & Francis Group, LLC, (2016).
2. P. Senthilkumaran, “Singularities in Physics and Engineering. Properties, methods, and applications”. IOP Publishing Ltd, (2018).
3. M.S. Soskin, M.V. Vasnetsov. “Singular optics”. Progress in optics, **42**(4), 219-276, (2001).
4. I.I. Mokhun, “Introduction to linear singular optics”, Chapter 1 in the book [Optical correlation techniques and applications], edited by O.V.Angelsky, SPIE press, Bellingham, Washington, USA, pp. 1-132, (2007).
5. A. Bekshaev, K. Bliokh, M. Soskin. “Internal flows and energy circulation in light beams”. J. Opt. **13** (5), 053001, (2011).
6. A. Bekshaev, M. Soskin. “Transverse energy flows in vectorial fields of paraxial beams with singularities”. Opt. commun., **271**, pp. 332-348, (2007).
7. M.V. Berry, M.R. Dennis. “Stream function for optical energy flow” J. Opt., **13**, 06400415, (2011).
8. J.H. Poynting “The wave motion of a revolving shaft, and a suggestion as to the angular momentum in a beam of circularly polarised light”, Proc. Roy. Soc. London A **82**, pp. 560–567, (1909).
9. R.A. Beth “Mechanical detection and measurement of the angular momentum of light”, Phys. Rev. **50**, pp. 115–125, (1936).
10. L. Allen, M.J. Padgett, and M. Babiker. “The orbital angular momentum of light,” E.Wolf, Progress in optics **XXXIX**, Elsevier Science B.V., (1999).
11. D. McGloin. “Optical tweezers: 20 years on”. Phil. Trans. R. Soc. A. **364**, pp. 3521–3537, (2006).
12. M.J. Lang, S.M. Block. “Resource Letter: LBOT-1: Laser-based optical tweezers”. Am. J. Phys. **71**, pp. 201–215, (2003)
13. I. Mokhun, R. Khrobatin. “Shift of application point of angular momentum in the area of elementary polarization singularity”. J. Opt. A: Pure Appl. Opt. **10,** 064015, (2008).
14. I. Mokhun, A. Mokhun, Ju. Viktorovskaya. “Singularities of Poynting vector and the structure of optical fields”. UJPO, **7**, pp. 129–141, (2006).
15. I. Mokhun, Y. Galushko, Y. Viktorovskaya, M. Karabchyivskiy, A. Bekshaev. “Transformations of the transverse Poynting vector distribution upon diffraction of a circularly polarized paraxial beam”. J. Opt. Soc. Am. A, **41**(3), pp. 382–391, (2024).
16. J. Olmos-Trigo, D.R. Abujetas, C. Sanz-Fernández, et al., “Unveiling dipolar spectral regimes of large dielectric Mie spheres from helicity conservation,” Phys. Rev. Res. **2**, 043021, (2020).
17. I. Mokhun, A. Arkhelyuk, Yu. Galushko, Ye. Kharitonova, Ju. Viktorovskaya, “Experimental analysis of the Poynting vector characteristics”, Appl. Opt., **51**, pp.C158-C162, (2012).
18. I. Mokhun. “Validity of running criterion”. Proc. SPIE., **9809**, 980904, (2015).
19. J. Hartmann. "Bemerkungen uber den Bau und die Justirung von Spektrographen," Z. Instrumentenkd., **20**, 47 (1900).
20. M. Born and E. Wolf. “Principles of optics”, sixth edition, Oxford: Pergamon, (1980).
21. N.R. Heckenberg, R. McDuff, C.P Smith. and A.G. White. “Generation of optical singularities by computer-generated holograms”, Opt. Lett., **17**, pp. 221-223, (1992).
22. I.V. Basisty, M.S. Soskin, M.V. Vasnetsov. “Optical wavefront dislocations and their properties”, Opt. Comm., **119**, pp.604-612, (1995).